\documentclass[
reprint,
superscriptaddress,
 amsmath,amssymb,
 aps,
 prb,
]{revtex4-1}
\usepackage{graphicx}
\usepackage{dcolumn}
\usepackage{epstopdf}
\usepackage{bm}
\usepackage{hyperref}
\usepackage{amssymb}
\usepackage{amsmath}
\usepackage{amsfonts}
\usepackage[english]{babel}
\begin{document}

\preprint{APS/123-QED}

\title{Effect of a skin-deep surface zone on formation of two-dimensional electron gas at a semiconductor surface}

\author{Natalia Olszowska}
\author{Jakub Lis}
\author{Piotr Ciocho\'{n}}
\affiliation{
Faculty of Physics, Astronomy, and Applied Computer Science, Jagiellonian University, {\L}ojasiewicza 11, 30-348 Krak{\'o}w, Poland
}
\author{{\L}ukasz Walczak}
\author{Enrique G. Michel}
\affiliation{
Dpto. de F\'{\i}sica la Materia Condensada and Condensed Matter Physics Center (IFIMAC), Universidad Aut\'{o}noma de Madrid, 28049 Madrid, Spain
}
\author{Jacek J. Ko{\l}odziej}
\email{jj.kolodziej@uj.edu.pl}
\affiliation{
Faculty of Physics, Astronomy, and Applied Computer Science, Jagiellonian University, {\L}ojasiewicza 11, 30-348 Krak{\'o}w, Poland
}

\date{\today}

\begin{abstract}

Two dimensional electron gases (2DEGs) at surfaces and interfaces of semiconductors are described straightforwardly with a 1D self-consistent Poisson-Schr\"{o}dinger scheme. However, their band energies have not been modeled correctly in this way. Using angle-resolved photoelectron spectroscopy we study the band structures of 2DEGs formed at sulfur-passivated surfaces of InAs(001) as a model system. Electronic properties of these surfaces are tuned by changing the S coverage, while keeping a high-quality interface, free of defects and with a constant doping density. In contrast to earlier studies we show that the Poisson-Schr\"{o}dinger scheme predicts the 2DEG bands energies correctly but it is indispensable to take into account the existence of the physical surface. The surface substantially  influences the band energies beyond simple electrostatics, by setting nontrivial boundary conditions for 2DEG wavefunctions. 

\begin{description}
\item[68.35.bg, 73.20.At, 74.20.Pq, 79.60.Bm]
\end{description}
\end{abstract}

\pacs{Valid PACS appear here}
\maketitle


\section{\label{sec:Introduction}Introduction}

Two-dimensional electron gases (2DEGs) occuring at surfaces of semiconductors  have been studied since many years due to their rich phenomenology and extreme technological relevance  \cite{NoguchiPRL91, OlssonPRL96, KingPRL2010,  ColakerolPRL2006, PiperPRB2008, ZhangPRL2013,  MeewasanaNatMat2011, ColakerolSS2015, AristovPRB99, BettiPRB2001, KobayashiPRL115, WalkerADVMat27, SyroNatMat13}. The 2DEGs arise following subsurface confinement of conduction electrons caused by an electric field.  A characteristic quasi-2D surface electronic structure (a set of 2D subbands) is then observed. Depending on the semiconductor doping the phenomenon may take the form of the charge accumulation or the charge inversion layer. A few crystals host native charge accumulation/inversion layers at their surfaces, notably InAs, CdO, InN, In$_2$O$_3$, and SrTiO$_3$. In many other cases the layers can be intentionally engineered, using adsorbates. Analogous phenomena are found at many interfaces \cite{FowlerPRL66,ChangPRL111,Ando,Ohtomo,ChenNano,CancellieriNat,Schmickler}. 

Theoretical modeling of the 2DEGs builds upon the assumption that the surface electrostatic potential is screened by a degenerate electron gas residing in a subsurface potential well. This is formulated as a 1D self-consistent Poisson-Schr\"{o}dinger problem. The problem has been solved  iteratively \cite{DukePRB67, BaraffPRB71, BaraffPRB72,GaoJAP2013} and also using the modified Thomas-Fermi approximation (MTFA) \cite{UbenseePSS82, UbenseePSS88, UbenseePRB89, KingPRB2008}. These two strategies have been found equivalent \cite{UbenseePSS88, EhlersPRB86}. However, angle-resolved photoelectron spectroscopy (ARPES) experimental tests of the 2DEG band structure have shown that the subband energies are not described well by the models \cite{KingPRL2010, ZhangPRL2013, MeewasanaNatMat2011}. It has been then proposed that taking into account many-body interactions could resolve the problem \cite{KingPRL2010}. However, these interactions are known to be negligible \cite{BaraffPRB72} at least in some of the problematic 2DEG systems. Thus today, nearly 50 years after the first experimental evidence of 2DEG  \cite{FowlerPRL66}, there is still no complete understanding of the 2DEG systems. 

In the present paper we revisit the problem by combining experimental and theoretical studies. Experimental ARPES results and theoretical calculations within the schemes mentioned above \cite{DukePRB67, BaraffPRB71, BaraffPRB72, UbenseePSS82, UbenseePSS88, UbenseePRB89, KingPRB2008}, are brought into harmony after considering adequate boundary conditions for the 2DEG wave functions at the surface. These boundary conditions are traced back to the nontrivial potential interpolation between the crystal and vacuum

\section{\label{exp}Experimental methods}
We focus on InAs, as it is probably the best studied material showing native accumulation layers~\cite{NoguchiPRL91, OlssonPRL96, KingPRL2010, BettiPRB2001, AristovPRB99, LoweSS2003,  WeberAPL97, TomaszewskaSS2015}, having as well a large technical application potential \cite{Rehm,Chang,Yamaguchi}.
The band bending at InAs surface depends on its orientation and reconstruction as well as on adsorbates  \cite{OlssonPRL96, TomaszewskaSS2015, AristovPRB99, AristovJVST94, BettiPRB2001,LoweSS2003, LoweSS20032, LoweJCG2002}. We use sulfur treatment on the (001) surface in order to control the bending \cite{note2}. 

Our samples are nominally undoped, n-type InAs(001) wafers with a donor concentration of  $3\times10^{16}$ cm$^{-3}$. Sputtering of the surfaces is avoided, as it produces many electrically active defects in the subsurface region \cite{TomaszewskaSS2015, BellPRB96}. The samples are treated with a HCl-iPA solution in 5N nitrogen  atmosphere, rinsed by iPA, blown dry, and transferred to ultra-high vacuum without being exposed to atmospheric air, then annealed to prepare In and As terminated surfaces \cite{TomaszewskaSS2015, OlszowskaSS2015}. A S$_{2}$ beam is generated using an electrochemical cell \cite{Heegemann,Sulfurtune}. The S-adsorbed surfaces are annealed in steps to increasingly higher temperature and their reconstructions are monitored using LEED. Most of ARPES studies are done using He I$\alpha$ radiation, at 80 K. Variable photon-energy ARPES spectra are acquired at APE beamline at Elettra storage ring in Trieste (Italy). 

We study a few differently reconstructed surfaces, prepared with S adsorption on both In and As-terminated InAs(001). Little is known about their atomic structures \cite{Tsukamoto, FukudaPRB97, LoweSS20032, Katayama} but it is not crucial for our discussion, as only values of the total band bending explicitly enter the model under consideration. A very important observation is that, due to the relatively low processing temperatures \cite{Schillmann,Kato69}, the doping density in the subsurface region remains unchanged for all samples studied and equals that of the bulk. Thus unnecessary complications such as uncertain $\Delta$ (see below for the definition) and unbound-charge screening effects are eliminated on the experimental level.

\section{\label{res}Experimental results}
\begin{figure}[h!]
\includegraphics[width=8.5cm]{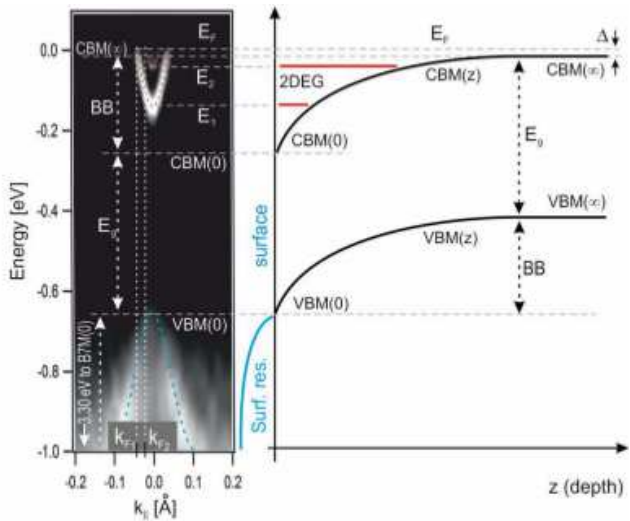} 
\caption{\label{pasma1}(color online) ARPES image of the electronic structure of exemplary InAs surface including a 2DEG and the corresponding band diagram.}
\end{figure}

The band diagram defining the parameters and conventions used throughout the paper is shown in Fig. \ref{pasma1}. $E_F$ is the Fermi level for which we assign the zero energy. The bulk electronic structure is represented by a depth dependent valence band maximum VBM($z$) and conduction band minimum CBM($z$). The band gap ($E_{g}$) is the difference between the two. At the temperatures relevant to our experiment ($\sim$80 K)  $E_{g}$ is equal to 0.40 eV \cite{Fang1990}. Far from the surface, the $E_{F}$ to CBM($\infty$) distance is denoted as $\Delta$. $\Delta$ is calculated using standard formulas \cite{Grundmann} and it is small ($\sim$10 meV) for our samples. The total band bending is obtained as: $BB=-VBM(0)-E_{g}-\Delta$. Throughout this paper the variable $z \geq 0$ denotes the depth beneath the surface
located at $z = 0$.

\begin{figure}[h!]
\includegraphics[width=8.5cm]{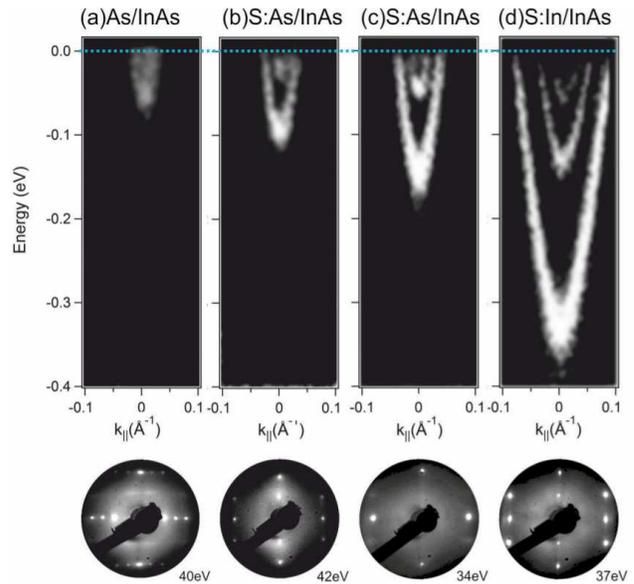}
\caption{\label{subbands} ARPES photocurrent maps reflecting 2DEG bands on InAs(001) surfaces, along $\overline{\Gamma} \overline{J'}$ direction; (a) on $c(2\times8)-2\times4$ As-rich surface; (b),(c) on sulfur-passivated As-rich  $2\times1$ and $1\times1$ reconstructed surfaces, respectively ; (d) on sulfur-passivated In-rich surface reconstructed $2\times1$. LEEDs for the surfaces are shown in the lower row.  }
\end{figure}

Our study shows that the 2DEG subbands on InAs are not well resolved (as in many earlier reports, see Ref. \onlinecite{note3}) if the sample quality is not excellent - see the exemplary spectrum in Fig. \ref{subbands}(a). However, we limit our discussion to examples where the subbands are seen clearly - {\it cf.} Fig. \ref{subbands}(b-d). The subbands are found isotropic, so there is no need to consider their azimuthal orientation.  

\begin{figure}[h!]
\includegraphics[width=8.5cm]{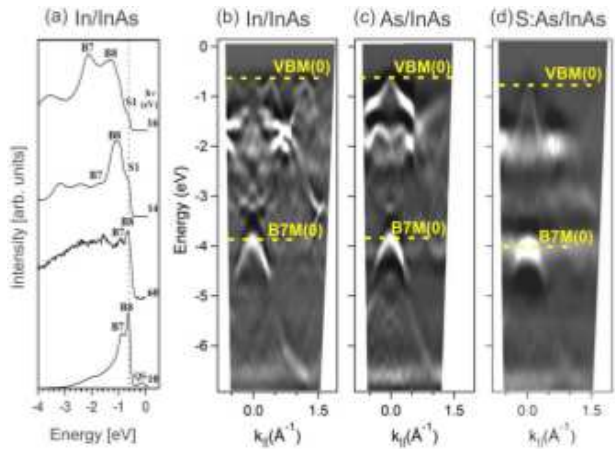}
\caption{ \label{vbm-graphs}(color online) (a) Valence band of In-rich $c(8\times2)-4\times2$ InAs(001) investigated using variable photon energy ARPES at normal emission (second derivative of the   photocurrent). Bulk (B7, B8) and surface (S1) related features are indicated. QS corresponds to emission from the 2DEG. (b),(c),(d) Valence band structures along $\overline{\Gamma} \overline{J'}$ direction imaged with ARPES: correspondingly for $c(8\times2)-4\times2$ In-rich surface, $c(2\times8)-2\times4$ As-rich surface, $1\times1$ S passivated As-rich surface. Lines denoting VBM(0) are drawn 3.30 eV above the lines denoting B7M(0).}
\end{figure}

In order to determine the BB value we notice that the VBM in InAs corresponds to the $\Gamma_8 $ level, which is the top of the B8 bulk band \cite{Chelikowski}. When measured with normal-emission, variable excitation energy ARPES, the B8 band reaches maximal value at 10 eV photon energy and again at 60 eV [see Fig. \ref{vbm-graphs}(a) and Appendix A], meaning that the bulk $\Gamma$ point is probed for these energies. We also notice that, for the In-terminated (001)InAs surface, there exists a nondispersing surface resonance (S1), clearly seen at photon energies of  14 and 16 eV, when it is separated from bulk bands. As seen in Fig. 3(a) VBM(0) and S1 coincide (at -0.55 eV). Thus, for the In-terminated (001)InAs surface one may also find the VBM(0) simply by studying the onset of the valence band. 
In order to determine the BB using He I$\alpha$ excited ARPES we find the surface independent reference, that is the apparent B7 band maximum [B7M(0)] - see  Fig. \ref{vbm-graphs}(b). Based on the detailed studies of the clean In-rich InAs surface we find the difference in energy between the VBM(0) and B7M(0) as 3.30 eV. Then the BB values are found using the formula: $BB=-B7M(0) - 3.30 eV - E_{g}-\Delta$  [see also  Figs. \ref{vbm-graphs} (c) and (d)]. As seen in Fig. \ref{vbm-graphs}, surface resonances are present at the $\overline{\Gamma}$ point, aligned with the VBM(0) to within accuracy of our measurements, both for clean and S-passivated InAs(001) surfaces. Similar results have been obtained before \cite{TomaszewskaSS2015, KingPRB2008, WalczakSS2013}.

\begin{table}
\center
\caption{\label{table1} Measured 2DEG properties for clean and S passivated InAs surfaces. See Fig. \ref{pasma1} for definitions of the parameters shown.  
The sheet electron density $n_{2D}$ is estimated by the Luttinger area: n$_{\bf 2D}=\sum_{i}{k_{F_{i}}^{2}}/{(2\pi)}$. 
 Uncertainties are: 50 meV for VBM(0) and BB, 5 meV for E$_{i}$, 0.005 \AA$^{-1}$ for $k_{F_{i}}$.}
\begin{ruledtabular}
\begin{tabular}{ l  c c  c c c }

& \multicolumn{2}{ c }{ on the In-rich surf.} & \multicolumn{3}{ c }{on the As-rich surface}\\
\colrule
\textbf{} & \textbf{clean} &  {\bf 2$\times$1(S)} & \textbf{clean} & {\bf 1$\times$1(S)} & {\bf 2$\times$1(S)}\\
\colrule
\\
{\bf VBM(0)} [eV]                       & -0.55 & -0.98  & -0.52 & -0.69  & -0.62\\
{\bf BB} [eV]                           & 0.15  & 0.58   & 0.12  & 0.29   & 0.22\\

{\bf E$_{\bf 1}$} [eV]                  & --    & -0.33  & --    & -0.15  & -0.09\\
{\bf E$_{\bf 2}$} [eV]                  & --    & -0.13  & --    & -0.04  & -0.02\\
{\bf E$_{\bf 3}$} [eV]                  & --    & -0.05  & --    & --     & -- \\

{\bf k$_{\bf F_{1}}$} [\r{A}$^{-1}$]    & --    & 0.090  & --    & 0.040   & 0.030\\
{\bf k$_{\bf F_{2}}$} [\r{A}$^{-1}$]    & --    & 0.050  & --    & 0.015   & 0.010\\
{\bf k$_{\bf F_{3}}$} [\r{A}$^{-1}$]    & --    & 0.015  & --    & --     & -- \\
{\bf n$_{\bf 2D}$} [10$^{12}$cm$^{-2}$] & --    & 17(2) & --     & 2.9(7) & 1.6(5) \\

\end{tabular}
\end{ruledtabular}

\end{table}

Based on  the spectra shown in Fig. \ref{subbands} one may extract a few numerical results including the energies corresponding to the subband minima ($E_i$) and Fermi wavevectors for the subbands ($k_{F_{i}}$)  - see also Fig. \ref{pasma1} for explanation. These results are given in Table \ref{table1}. 
Earlier studies [see for example Ref. \onlinecite{KingPRL2010}] have revealed that  MTFA  underestimates the binding energies corresponding to subband minima E$_i$. Our experimental data are similarly incoherent with the energy spectra calculated from the MTFA,  see the case  $\lambda \rightarrow \infty$ [i.e. arctan$(\lambda) >1$] in  Fig.~\ref{theory1}. 

\section{ \label{theory} Theoretical framework}
In order to explain such discrepancies King et al. \cite{KingPRL2010} proposed that strong many-body interactions within the accumulation layers (neglected when using MTFA) cause a giant renormalization of the InAs bandgap down to 0.1 eV, at the surface.  We find this statement questionable. Our experimental bands exhibit no features \cite{Ingle2005Hofmann2009} pointing out to strong many-body interactions. We also notice that, in narrow direct-gap III-V semiconductors, for typical electron densities found in accumulation layers ($10^{18} - 10^{19}/cm^{3}$), the many-body interactions do not impact the band-gap substantially. This follows from the fundamental theory of the electron gas - see Ref. \onlinecite{BaraffPRB72}. These theoretical expectations are firmly confirmed experimentally - the bandgap renormalization is measured for degeneratively n-doped GaAs in several studies - see Ref. \onlinecite{Luo2002} and references therein. The renormalization is found not to exceed 100 meV, for doping concentrations $10^{18} - 10^{19}/cm^{3}$.  This is scaled  with  the factor $\sim$ 1/2 to represent InAs\cite{Palankovsky1999} but more than 3/4 of the effect is due to interactions of conduction electrons with ionized donors \cite{Jain1990}, non existent for the 2DEG case. Thus, the many-electron interactions, for the relevant densities, reduce the InAs bandgap only by $\sim$ 10 meV. While there are indications that the discussed effects increase when going from 3D to 2D systems, they still remain definitely insufficient \cite{plasmas}, i.e. not relevant in the first order for the InAs 2DEG band structure description. 

Having said this, we expect the one-electron Poisson-Schr\"{o}dinger calculation to be applicable to InAs. We follow the calculation scheme outlined in Ref.~\onlinecite{KingPRB2008} using the effective mass and envelope wave function approximation but we lift artifical surface boundary conditions imposed by MTFA. We discuss the calculation procedure in Appendix~\ref{App2} in more details. There are two steps in the calculations. First, the electrostatic potential $V$ is calculated within the band bending approximation from the one dimensional Poisson equation. Second, one-dimensional Schr{\"o}dinger equation is solved [the separated solutions in the dimensions parallel to the surface are left intact]
\begin{equation}
\left[-\frac{\hbar^{2}}{2m_{cb}}\frac{d^{2}}{dz^{2}}+V(z)\right]\psi=E\psi,
\end{equation}
where $m_{cb}$ stands for the effective mass in the conduction band. In general, the above equation on the semi-axis corresponds to a Hermitian (self-adjoint) operator if it acts on functions $\xi$  satisfying the following boundary condition at the origin~\cite{Bonneau}
\begin{equation}\label{condition}
\tfrac{d}{dz}\xi(0)=\lambda \xi(0),
\end{equation}
where $\lambda$ has dimension of inverse length and can have any value including infinity. 
We limit our consideration to $\lambda>0$ -  see Appendix~\ref{App3} for a justification. So far only the particular case with $\lambda=\infty$ corresponding to $\xi(0)=0$ has been considered~\cite{UbenseePRB89,KingPRB2008}, resulting in the basis of generalized wavefunctions (in the direction normal to the surface) ${sin(kz)}_{k>0}$. 

\begin{figure}[h!]
\includegraphics[width=8cm]{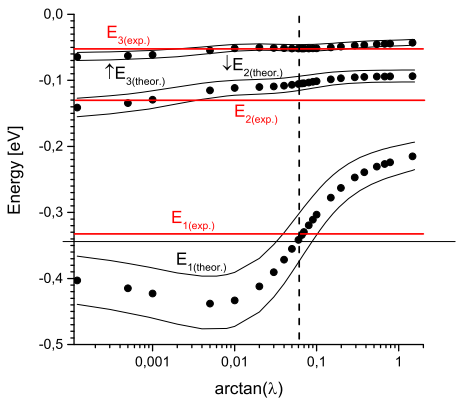}
\caption{ \label{theory1}(color online) Calculated dependencies of the theoretical 2DEG subband minima on the parameter $\lambda$ (points), for the case of BB=0.58 eV. Thin solid lines show the binding energies for BB $\pm$ 50 meV, which reflects the experimental uncertainty. Corresponding experimental energies are indicated by horizontal lines (red online). For arctan$(\lambda)>1$ MTFA solutions are reached. The vertical line indicates the $\lambda$ value for which the theoretical and experimental binding energies $E_i$ match the most closely.}
\end{figure}

We treat $\lambda$ as a parameter to be fit to the data.  As shown in  Fig. \ref{theory1}, the subband energies $E_i$ heavily depend on $\lambda$ and it is possible to nearly match the calculated and the experimental energies $E_{i}$ by selecting $\lambda$. Complete sets of energies $E_{i}$ evaluated within this scheme are shown in Table \ref{table2}. They are  close to the experimental values given in Table \ref{table1}. 

\begin{table}
\center
\caption{\label{table2} Calculated energies of the subbands minima for 2DEG corresponding to the experimentally investigated band bending and related values of  $\lambda$. }
\begin{ruledtabular}
\begin{tabular}{ l  c   c c c }

\textbf{BB} &  0.58 eV  &0.29 eV &0.22 eV\\
\colrule
\\
{\bf E$_{\bf 1}$} [eV]                     & -0.340  &-0.150  &-0.090\\
{\bf E$_{\bf 2}$} [eV]                     & -0.130  &-0.045  &-0.030\\
{\bf E$_{\bf 3}$} [eV]                     & -0.050  &        &       \\

{\bf $\lambda  $} [\AA$^{-1}$]         & 0.05  & 0.05   &0.09\\

\end{tabular}
\end{ruledtabular}

\end{table}

\section{\label{discuss}Discussion} 
Fig.~\ref{waveffig} schematically illustrates how $\lambda$ impacts ground-state solutions $\psi$. For $\lambda=0$ the maximum of the wavefunction is located at the origin, corresponding to a large negative potential energy $\langle \psi|V|\psi\rangle$. As $\lambda$ grows, the maximum appears at some distance from $z=0$, decreasing the binding energies of the 2DEG electrons. Another important factor is the kinetic energy, which measures the variation of the wavefunction amplitude and hence it is larger for $\lambda=\infty$ than for $\lambda=0$.  The  values of $\lambda$ given in Table~\ref{table2} are relatively small and the resulting wavefunction amplitudes at the surface are significant. In agreement with this discussion we have shown recently, using ARPES \cite{TomaszewskaSS2015}, that on a clean InAs(001) surface, 2DEG states acquire the periodicity of the surface reconstruction. This is an experimental evidence that the 2DEG wavefunctions amplitudes may be large at the very surface.

\begin{figure}
\begin{center}
\includegraphics[scale=0.55]{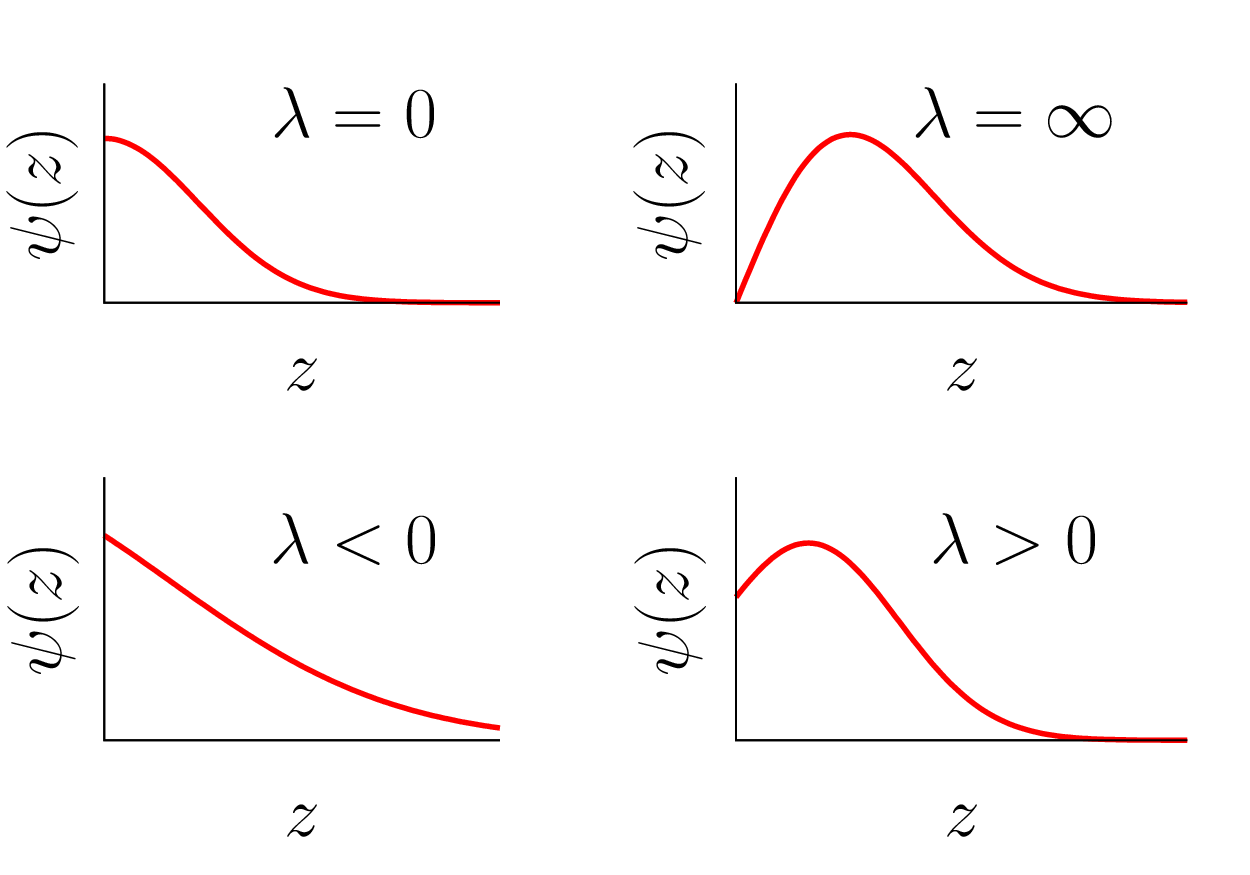}
\caption{\label{waveffig}(color online) Schematic wavefunctions of the bound state for different values of $\lambda$, as indicated.}
\end{center}
\end{figure}

Relation~(\ref{condition}) is in fact a generalization of the so-far considered model. To see this, we write the basis of generalized eigenfunctions as $\cos\left(kz+\phi(k)\right)$. The phase factor $\phi$ satisfies 
\begin{equation}\label{war1}
\cos\left(\phi(k)\right)=-sign(\lambda)\frac{k}{\sqrt{k^{2}+\lambda^{2}}}.
\end{equation}
For small wave vectors ($k\to 0$) the phase $\phi$ approaches $-\pi/2$, so $\cos(kz +\phi(k)) \approx \sin( kz )$; $\lambda=0$ is the only exception from the rule. 

In the envelope wavefunction approximation the trigonometric functions on the semi-axis can be regarded as coming from the interference between the incoming and outgoing waves 
\begin{equation}\label{phi}
\frac{1}{2}\left(e^{i k z}+e^{-i k z+2i \phi}\right)= e^{i\phi}\cos(kz-\phi)
\end{equation}
with the reflection coefficient equal to unity. The phase $\phi$ is the only remnant of the of the reflection caused by a non-trivial potential far from the surface. The condition  used so-far is equivalent to $\phi=\pm\pi$ and its rationale is given in the literature: ``As the characteristic penetration length of the wave functions into the vacuum is very short, much shorter than the variation of carrier density in the space-charge region, it is a good approximation to impose the boundary condition that the wave function is equal to zero at the surface and thus that the carrier concentration tends smoothly to zero at the surface''~\cite{KingPRB2008}. No doubt, the electron wavefunction dies-off outside the crystal. But, this does not mean that the the generalized wavefunctions have to be $\sin(\cdot)$ functions. This is true only if the surface is modeled as a featureless, infinite potential step, see Appendix~\ref{App3}.

The envelope wavefunction is an effective description of phenomena on large scales compared to the lattice constant. Thus, there is no unambiguous method to put the surface plane into the system. Furthermore, there is an about 1 nm wide zone at the surface corresponding to the reconstructed layers. In this zone the assumptions of the envelope approximation do not hold. So, the realistic model of 2DEG should include three regions:  the vacuum region with negligible wavefunction density, the crystal bulk where free-electron approximation works well and a transition zone where the potential interpolates between its vacuum and bulk values. We solve the electrostatic problem in the crystal bulk leaving  $\lambda$ as an effective parameter accounting for the transition zone properties. In Appendix~\ref{App3} we discuss simple models illustrating these ideas. 

\section{\label{conc}Conclusions}
We have presented coherent experimental and theoretical studies on 2DEG band structure for a few differently reconstructed InAs(001) surfaces, treated as a model system. The surfaces has been carefully chosen to avoid experimental problems that are often present for the kind of the spectroscopic studies done. Surface band bending values have been cautiously evaluated. Thus, the obtained experimental material provides a robust test of the applied theoretical concepts.
We find that a simple one-electron Poisson-Schr\"{o}dinger model explains the 2DEG band energies well, provided that the surface model is not oversimplified, i.e., proper boundary conditions on the 2DEG wavefunctions are imposed. We show that the conditions effectively describe the ``skin-deep surface zone'' (or the physical surface) and correspond to a nontrivial surface potential. Neglecting the ``skin-deep surface zone'' effect, what has been overlooked so far, leads to heavily underestimated 2DEG band energies. For InAs we find this effect solely being able to reconcile the measured and calculated energy spectra. Therefore we think that many-body corrections may be not taken into account  to first order, in the context of surface 2DEG in InAs. Whether they may be more pronounced for other materials, should be considered  in further studies.  The ``skin-deep surface zone'' effect must be, {\it in principle}, present for any 2DEG system, so that our findings impact also the understanding and modeling of two-dimensional electron gases existing at surfaces and interfaces of other semiconductors and oxides.

\begin{acknowledgments}
We acknowledge financial support by Polish NCN (contract 2011/03/B/ST3/02070). The research was carried out with the equipment purchased thanks to European Regional Development Fund in the framework of the Polish Innovation Economy Operational Program (contract no. POIG.02.01.00-12-023/08). L.W. and E.G.M. acknowledge financial support by MINECO (grant MAT2014-52477-C5-5-P). J.L. wishes to thank prof. K. Ro{\'s}ciszewski for instructive  discussion.
\end{acknowledgments}

\appendix
\section{\label{App1} Determination of VBM(0)}
This Appendix is intended to assist the interpretation of the spectra  shown in Fig. 3a.  Four parts of the Fig.  \ref{suppl1} show angle-resolved photocurrent maps measured using different photon energies. The two top maps correspond to 10 and 60 eV photon energies for which the bulk $\Gamma$ point is probed. For these maps the valence band region is dominated by bulk bands B8 and B7 and the VBM may be directly read as the energy of the B8 band at $k_{||}=0$. In contrast, the two bottom maps correspond to 14 and 16 eV photon energies and illustrate the situation when the B8 band is seen away from the VBM. In these cases the valence band region is dominated by surface resonances, including the indicated S1 coinciding, at $k_{||}=0$, with the VBM.

\begin{figure}[h!]
\setcounter{figure}{0}
\makeatletter 
\renewcommand{\thefigure}{A\@arabic\c@figure}
\makeatother

\includegraphics[width=8cm]{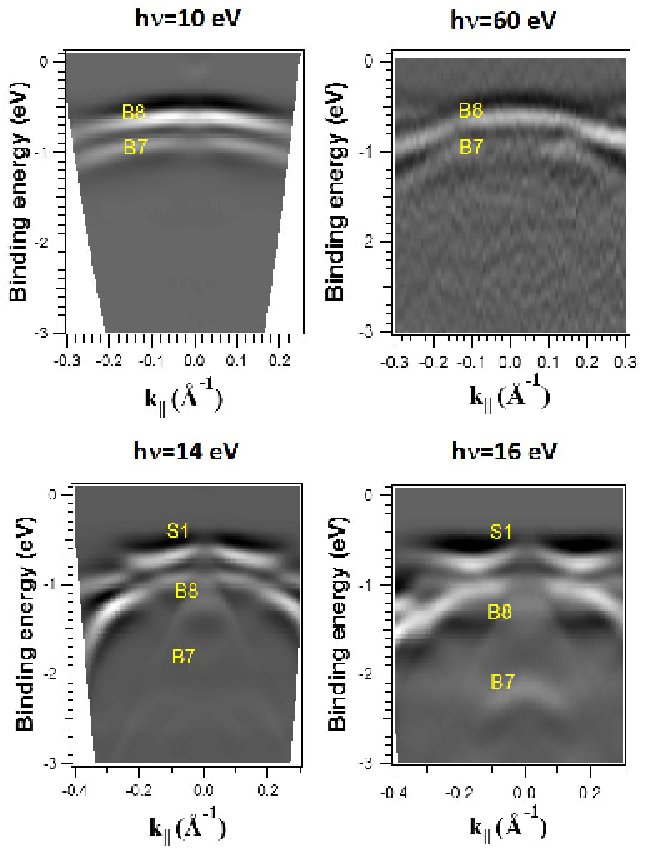}
\caption{ \label{suppl1}(color online)  ARPES photocurrent maps (second derivative) along $\Gamma J'$ for 10 eV and 60 eV (top) and for 14 eV and 16 eV (bottom) photon energy. The maps are collected for the InAs crystal terminated with clean In-rich (001) surface, reconstructed c(8x2)-4x2. Energy is measured relative to the Fermi level.}
\end{figure}

\section{\label{App2}Calculation scheme}
Here we outline the calculation procedure in detail. As the numerical procedures involved are not recourse-demanding we take into account the following bands: light holes, heavy holes, conduction band and  donor density (this is done for the universality of the model while for the investigated samples  the hole bands as well as the donor density factor could be neglected). The conduction band minimum corresponds to the zero energy. The charge density accumulated in the conduction band reads 
\begin{equation}
n_{cb}=\frac{1}{\pi^{2}}\int_{0}^{\infty}\frac{k^{2} \ dk}{1+\exp{\beta\left[E_{cb}(k^{2})-E_{F}+V(z)\right]}},
\end{equation}
where $\beta=1/k_{B}T$ with $k_{B}$ denoting the Boltzmann constant, $T$ the temperature and $V(z)$ is the electrostatic potential multiplied by the electron charge. We use the $\mathbf{k\cdot p}$ relation for the kinetic energy of the electrons
\begin{equation}
E_{cb}(k^{2})=\frac{\hbar^{2}}{2m_{e}}k^{2}+\frac{E_{g}}{2}\left(\sqrt{1+4 P^{2}k^{2}}-1\right),
\end{equation}
where $m_{e}$ stands for the free electron mass. $E_{g}$ is calculated from the Varshni parameterization \cite{KingPRB2008}
\begin{equation}
E_{g}=0.415 - \frac{2.76 \cdot 10^{-4} T^{2}}{83 + T},
\end{equation}
where $T$ is given in Kelvin and the energy in eV, and 
\begin{equation}
P^{2}=\frac{3\hbar^{2}}{2m_{e}}\left(\frac{m_{e}}{m_{cb}}-1\right)\frac{E_{g}+\delta}{(3 E_{g}+2\delta)E_{g}}.
\end{equation}
$\delta$ stands for the spin-orbit coupling, here $\delta=0.381$~eV. $m_{cb}=0.024m_{e}$ denotes  the electron effective mass in the conduction band. The formula for hole density reads
\begin{equation}
p_{i}=\frac{1}{\pi^{2}}\int_{0}^{\infty}\frac{k^{2} \ dk}{1+\exp{\beta\left(E_{g}+\tfrac{\hbar^{2}}{2m_{i}}k^{2}+E_{F}-V(z)\right)}},
\end{equation}
where $i\in\{lh, hh\}$ stands for light and heavy holes, respectively. The numerical values of the hole masses are $0.021m_{e}$ and $0.41m_{e}$, respectively. The density of (positively) ionized donors in non-degenerate semiconductors reads
\begin{equation}
\tilde{p}_{d}=\frac{N_{B}}{1+2\exp{\beta\left(E_{F}+E_{B}-V(z)\right)}},
\end{equation}
where the considered donor density is $N_{B}=3\cdot 10^{16} \ cm^{-3}$ and the energy of shallow donors is denoted with $E_{B}$. However, the average distance between the doping atoms is $N_{B}^{-1/3}=32$~nm, while the Bohr radius of the hydrogenic shallow donor states yields $33$~nm. So, at this density the material is rather a poor metal \cite{book}, and a full donor ionization should be assumed,
\begin{equation}
p_{d}=N_{B}.
\end{equation}
The electron neutrality condition for $V=0$, 
\begin{equation}
p_{lh}+p_{hh}+p_{d}-n_{cb}=0,
\end{equation}
sets the Fermi level. It is found 14 meV above the conduction band minimum.

Having set the Fermi energy, we can solve the Poisson equation
\begin{equation}\label{poisson}
\frac{d^{2}}{dz^{2}}V(z)=\frac{e_{0}^{2}}{\varepsilon_{0}\varepsilon_{b}}\left[ p_{hh}(z)+p_{lh}(z)+p_{d}(z)-n_{cb}(z)\right],
\end{equation}
where $\varepsilon_{0}$ and $\varepsilon_{b}$ stand for the vacuum dielectric constant  and InAs static dielectric constant, respectively, and $e_{0}$ for the electron charge. We require that  $V(0)$ be equal to the measured band bending and that $V(z)$ vanish in the bulk ($z\to\infty$). Hence, the derivative of $V(z)$ at the surface, which is interpreted as the surface charge, is not a free parameter. Now, we can solve the Schr{\"o}dinger equation: 
\begin{equation}
\left[E_{cb}\left(-\frac{d^{2}}{dz^{2}}\right)+V(z)\right]\psi=E\psi,
\end{equation}
where $E_{cb}\left(-\tfrac{d^{2}}{dz^{2}}\right)$ accounts for the non-parabolic dispersion relation. Various boundary conditions are assumed, as discussed in Sect.~\ref{theory}. Note that in Sect.~\ref{theory} we  discussed the parabolic dispersion relation while the actual calculations are performed for non-parabolic $E_{cb}(k)$. The rationale is that the theory of Hamiltonian operators  and related self-adjoint extensions on the semi-axis has been formulated for the Laplace operator. The status of the non-parabolic operators is not clear and our arguments loose their mathematical rigor. However,  they seem to be physically reasonable as discussed in Sect.~\ref{discuss} and below.

The above described scheme is not fully self-consistent. When calculating potential~(\ref{poisson}),  the plane wave approximation is assumed. This potential is subsequently used to pick out the correct boundary condition. Numerical investigations showed that   the choice of  boundary condition is not decisive for the potential. It is $\tfrac{d}{dz}V(0)$ which is sensitive to $\lambda$, making us reluctant to attribute to it the  strict physical meaning of the surface charge density.

\section{\label{App3}Tentative interpretation of $\lambda$ }
We  argue above that the non-trivial boundary condition imposed on the wavefunctions are due to the thin interpolation zone between the crystal and vacuum. To be more specific, we discuss below three simple models and show how these intuitions can be quantified. Here, to avoid technical complications we consider the parabolic dispersion. \\
We begin with the  well known step potential 
\begin{equation}
\Theta(z)=\left\{ \begin{array}{lr} 
\frac{\hbar^{2}}{2 m}\Theta_{0} & z<0,\\
0&0>z.\\
\end{array} 
\right.
\end{equation}
with $\Theta_{0}>0$. The  Schr\"odinger equation has the form
\begin{equation}\label{schrod}
\left(-\frac{\hbar^{2}}{2 m}\frac{d^{2}}{dz^{2}}+\Theta(z)-\frac{\hbar^{2}}{2 m}k^{2}\right)\psi(z)=0,
\end{equation}
where $\tfrac{\hbar^{2}}{2 m}k^{2}$ denotes the energy. The solutions read
\begin{equation}
\psi(z)=\left\{ \begin{array}{lr} 
\left(\sin\phi_{1}(k)\right) \exp{\sqrt{\Theta_{0}-k^{2}}z} & z<0,\\
\sin{\left(k z+\phi_{1}(k)\right)} &z>0.\\
\end{array} 
\right.
\end{equation}
The phase 
\begin{equation}
\tan \phi_{1}(k)=\frac{k}{\sqrt{\Theta_{0}-k^2}}.
\end{equation}
ensures equality of the  derivatives calculated at $0^{+}$ and $0^{-}$. For small $k$ the above equation simplifies to
\begin{equation}
\phi_{1}(k)=\frac{k}{\sqrt{\Theta_{0}}}.
\end{equation}
The non-trivial phase appears in the case of the step potential, it is negligible if $\Theta_{0}\gg k^{2}$, so it holds only in the case of the infinite barrier. The interpolating zone can be introduced by an additional step
\begin{equation}
\Theta(z)=\left\{ \begin{array}{lr} 
\infty & z<0,\\
\frac{\hbar^{2}}{2 m} U &0<z<z_{0},\\
0&z>z_{0}.\\
\end{array} 
\right.
\end{equation}
We solve the Schr\"odinger equation~(\ref{schrod}) for $0<k^{2}<U$ with the boundary condition $\psi(0)=0$. The solutions read
\begin{equation}\label{wavef1}
\psi_{+}(k;z)=\left\{ \begin{array}{cc}
a_{+}\sinh\left(\sqrt{U-k^{2}}z\right)&0<z<z_{0},\\
\cos\left(k z+\phi_{+} \right) & z>z_{0},\\
\end{array} \right.
\end{equation}
where $a_{+}$ and $\phi_{+}$ are parameters to be determined. The continuity of $\psi_{+}$ and its first derivative at $z_{0}$ result in the following relation for $\phi_{+}$
\begin{equation}\label{C8}
\sqrt{U-k^{2}} \coth\left(\sqrt{U-k^{2}}z_{0}\right)=-k \tan\left(kz_{0}+\phi_{+}\right),
\end{equation}
leading for small $k$ to the following formula
\begin{equation}\label{war11}
\cos\phi_{+}=-\frac{k}{\sqrt{k^{2}+k_{+}^{2}}},
\end{equation}
with 
\begin{equation}\label{war2}
k_{+}=\sqrt{U}\coth\sqrt{ U}z_{0},
\end{equation}
and
\begin{equation}\label{war3}
a_{+}=-\frac{k}{\sqrt{U}\cosh\sqrt{ U}z_{0}}.
\end{equation}

Eq.~(\ref{war11}) coincides with eq.~(\ref{war1}) upon identification $\lambda=k_{+}$ which allows drawing an analogy between low-energetic scattering from a non-trivial potential and the abstract condition~(\ref{condition}) for $\lambda>0$. 
The approximate value of the solutions is evident -- the resulting wavefunctions have a common part $\sinh \sqrt{U}z$ on the distance $0<z_{0}$. The matching condition is an approximate one and so are the resulting wavefunctions. But only this assumption allows  switching to the well-defined but simpler Hamiltonian. To this end we consider the potential
\begin{equation}\label{pot3}
V(z)=\left\{ \begin{array}{lr} 
\infty & z<0,\\
\frac{\hbar^{2}}{2 m} U+\nu(z) &0<z<z_{0},\\
\nu(z)&z>z_{0}.\\
\end{array} 
\right.
\end{equation}
The potential $\nu(z)$ varies slowly on distances much larger than $z_{0}$. Additionally,  we consider $U\gg \nu(z)$ for $(0,z_{0})$, so that the solutions~(\ref{wavef1}) are valid for $z<z_0$. Then  we arrive at   the equation 
\begin{equation}
\left(-\frac{\hbar^{2}}{2 m}\frac{d^{2}}{dz^{2}}+\nu(z)-\frac{\hbar^{2}}{2 m}k^{2}\right)\psi(z)=0
\end{equation}
for $z>z_{0}$ and with the boundary condition~$\psi'(z_{0})=k_{+}\psi(z_{0})$. As a consequence, we obtain an approximate spectrum of the problem~(\ref{pot3}), whose quality depends on the mutual relations between $U$ and $\nu$. Such a separation of problems for $z<z_{0}$ and $z>z_{0}$ is possible in the low energy region only. 

Similarly, one can treat the case with $U<0$. Then both $\lambda>0$ and $\lambda<0$ can appear. The latter case corresponds to the model with short-ranged potentials deep enough to produce a single bound state~\cite{Bonneau}.

 We limit our discussion to the low-energy sector. This is meaningful, if $k^{2}\ll U$, i.e., the energy of the wave is much lower than the barrier height. In physical terms, the barrier corresponds to the material work function (4-5 eV) which  is at least ten times more than  energies encountered in accumulation layers. There is also a term $k z_{0}$ in ~(\ref{C8}) which is neglected to arrive at eq.~(\ref{war11})-(\ref{war3}). As such, the approximation works well if the surface potential range is small compared to the wavevectors $k$.  \\

The step functions are nonphysical. Now we will consider  a model with a smooth potential $W$ exploding to infinity for $z=0$. The model has an interesting property: the wavefunctions  vanish at the origin and, nevertheless,  they are very close to  $\cos(\cdot)$ functions in the region where $W$ is negligible. As such, the parameter $\lambda$ considered above and the behavior of the wavefunctions at the very surface are not correlated. We consider function $f$,
\begin{equation}
f(z)=1-\frac{1}{2}\left(e^{-z}+e^{-z^{2}}\right)
\end{equation}
which linearly approaches zero at the origin ($f(z)\sim z/2$ for $z\to0$) and tends to a non-zero value for large $z$ ($f(z)\to 1 $ for $z\to\infty$). Using this function we define the potential
\begin{equation}
W(z)=\frac{\hbar^{2}}{2 m f(z) }\frac{d^{2}f(z)}{d^{2}z}
\end{equation}
that  explodes close to the origin ($W(z)\sim +z^{-1}$ for $z\to 0$) and quickly vanishes for large $z$, see Fig.~\ref{resonance}. Trivially, by plugging  $k=0$ and $\psi=f$ into   Schr{\"o}dinger equation 
\begin{equation}
\left[-\frac{\hbar^{2}}{2m}\frac{d^{2}}{dz^{2}}+W(z)-\frac{\hbar^{2}}{2m}k^{2}\right]\psi(z)=0
\end{equation}
one can check that $f$ is a bound, zero-energy solution. It is called  zero or threshold resonance~\cite{simon}. The relevance of $f$ lies in the fact that solutions with the same boundary condition [$\psi(0)=0$] approach (pointwise) $f(z)$ in the limit $k\to 0$~\cite{lis}. So, the low energy solutions are nearly constant in the region where $f$ is constant and, in this region, appear as $\cos(\cdot)$ functions, see Fig.~(\ref{resonance}) for a schematic explanation. As such, even $\lambda=0$ can concur with the finite penetration length. 
\begin{figure}
\setcounter{figure}{0}
\makeatletter 
\renewcommand{\thefigure}{C\@arabic\c@figure}
\makeatother
	\begin{center}
		\includegraphics[scale=0.45]{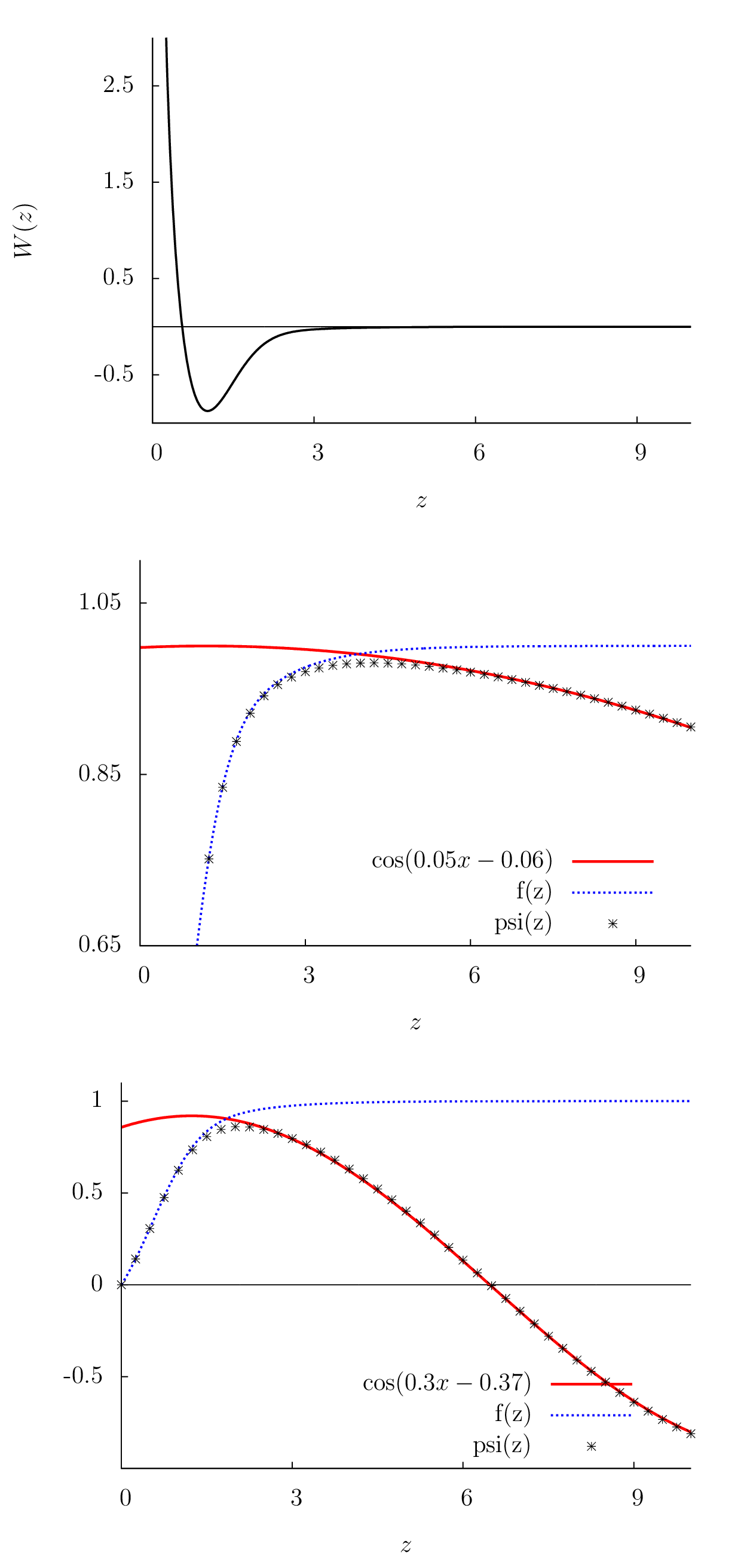}
		\caption{(color online) The threshold potential described in the text (A). Numerical solutions $\psi(z)$ for $k=0.05$ -- B, and $k=0.3$ -- C compared with fitted trigonometric functions. The smaller $k$, the smaller the resulting phase according to general arguments given in the text. Note that the smaller $k$ is considered, the longer the solutions $\psi$ stay close to  function $f$. For the linearity of the Schr{\"o}dinger equation we do not give amplidudes of the fitted functions.  \label{resonance}}
	\end{center}
\end{figure}

The models above show that the penetration length through a barrier and the phase shift are not tightly related and any value of $\lambda$ in eq.~(\ref{condition}) is admissible provided the existence of a short ranged potential interpolating between the crystal bulk and the vacuum.


\begin{thebibliography}{100}
\bibitem{note2} The sulfur treatment allows to keep ARPES images of 2DEG bands notably clear. The sulfur atoms are included in surface reconstructions to form a uniform (ordered) distribution of electrically active centers. This is in contrast to the case of adsorbed alkali atoms, (forming chains and islands on the surface), which cause spatially non-uniform band bending and a distorted subband structure. Moreover, the surface disorder has a detrimental effect on ARPES spectra via scattering photoelectrons in their final states.
\end{thebibliography}
\end{document}